\documentclass[12pt]{article}
\usepackage{epsfig}   
\usepackage{glas}   
\begin{document}
\begin{titlepage}{GLAS-PPE/1999--13}{September 1999}
\title{Hadronic Final States \\ in Deep Inelastic
Scattering at HERA}
\author{N.~H.~Brook\\
(on behalf of the H1 and ZEUS collaborations)}
\begin{abstract}
Results on the analysis of the hadronic final state 
in neutral current deep
inelastic scattering at HERA are presented; 
recent results on inclusive single particle distributions,
particle correlations and event shapes are highlighted.
\end{abstract}
\vfill
\conference{Invited talk at the Ringberg workshop:``New Trends in HERA Physics''\\
Tegernsee, Germany, 1999}
\end{titlepage}

\section{DIS kinematics}
The event kinematics of deep inelastic scattering, DIS,
are determined by the negative square of the four-momentum transfer at
the lepton vertex,
$Q^2\equiv-q^2$, and the Bjorken scaling variable, $x=Q^2/2P \cdot q$,
where $P$ is the four-momentum of the proton.
In the quark parton model (QPM),
the interacting quark from the proton carries the four-momentum $xP$.
The variable $y$, the fractional energy transfer to the proton in 
the proton rest
frame, is related to $x$ and $Q^2$ by $y\simeq Q^2/xs$, where $\sqrt s$
is
the positron-proton centre of mass energy.

Neutral current (NC) DIS occurs when an uncharged boson ($\gamma, Z^0$)
is exchanged between the lepton and proton.
In QPM there is a 1+1 parton configuration which consists of a single
struck quark and the proton remnant,
denoted by ``+1''.
At HERA energies there are significant higher-order
quantum chromodynamic (QCD) corrections:
to leading order in the strong coupling constant, $\alpha_{\rm s},$
these are QCD-Compton scattering (QCDC),
where a gluon is radiated by the
scattered quark and boson-gluon-fusion (BGF),
where the virtual boson and a gluon fuse to
form a quark-antiquark pair.
Both processes have 2+1 partons in the final
state.
There also exists calculations for the higher,
next-to-leading (NLO) processes.

A natural frame in which to study the dynamics of the hadronic final
state
in DIS is the Breit frame~\cite{feyn}.
In this frame, the exchanged
virtual boson ($\gamma^*$)
is completely space-like  and has a four-momentum
$q = (0,0,0,-Q=-2xP^{Breit})\equiv (E,~p_x,~p_y,~p_z)$,
where $P^{Breit}$ is the momentum of the proton in the Breit frame.
The particles produced in the 
interaction can be assigned to one of two regions:
the current region if
their $z$-momentum in the Breit frame is negative, and
the target region if their $z$-momentum is positive.
The main advantage of this
frame is that it gives a
maximal  separation of the incoming and outgoing partons
in the QPM. In this model
the maximum momentum a particle can have in the current region
is $Q/2,$ while in the target region the maximum is $\approx Q(1-x)/2x.$ 
In the Breit frame, unlike the hadronic centre of mass ($\gamma^*p$)
frame, 
the two regions
are asymmetric, particularly
at low $x,$ where the target region occupies most of 
the available phase space.

\section{Current Fragmentation Region}

The current region in the $ep$ Breit frame
is analogous to a single hemisphere of $e^+e^-$ annihilation.
In $e^+e^- \rightarrow q \bar q$ annihilation the two quarks are
produced
with equal and opposite momenta, $\pm \sqrt{s_{ee}}/2.$
The fragmentation of these quarks can be compared to
that of the quark struck from the
proton; this quark has an outgoing momentum $-Q/2$ in the Breit frame.
In the direction of this struck quark 
the scaled momentum spectra of the particles, expressed in terms of
$x_p = 2p^{Breit}/Q,$
are expected to have a
dependence on $Q$ similar to that observed in
$e^+e^-$~annihilation~\cite{eedis,anis,char} at energy $\sqrt{s_{ee}}=Q.$

\subsection{Evolution of $\boldmath \ln(1/x_p)$ Distributions}

Within the framework of the modified leading log approximation (MLLA)
there are predictions of how the higher order
moments of the parton momentum spectra should evolve 
with the energy scale~\cite{fongweb,dokevol}.
These parton level predictions in practice depend on
 two free parameters, a running 
strong coupling, governed by a QCD scale
$\Lambda,$ and an energy cut-off, $Q_0,$ 
below which the parton
evolution is truncated. 
In this case $\Lambda$ is an effective
scale parameter and is not to be identified with the standard QCD scale,
e.g. $\Lambda_{\rm \overline{MS}}$.
In particular, predictions can be made at
$Q_0 = \Lambda$ yielding the so-called limiting spectrum.
The hypothesis of
local parton hadron duality (LPHD)~\cite{LPHD},
which relates the observed hadron
distributions to the calculated parton distributions via a constant of
proportionality, is used in conjunction with the parton predictions
of the MLLA to allow the calculation to be directly compared to data.

\begin{figure}[htb!]
\centerline{\psfig{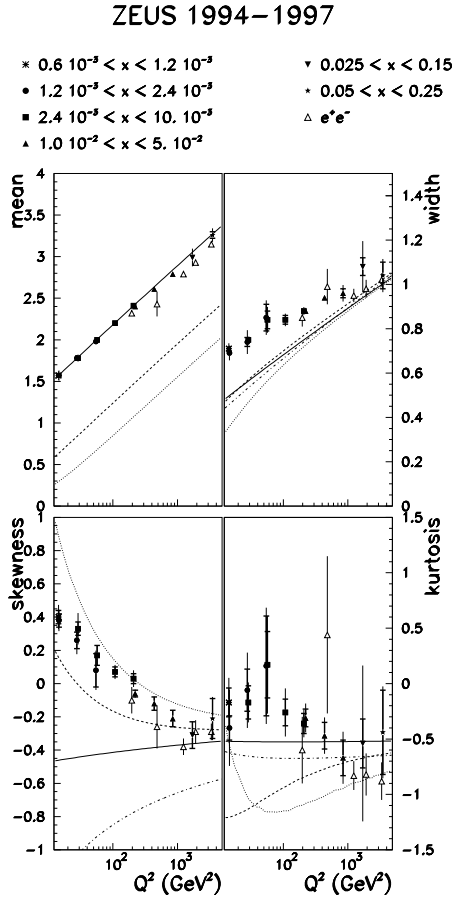}}
\caption{ Evolution of the mean, width, skewness and kurtosis of the
$\ln(1/x_p)$ distribution in the current fragmentation region with
$Q^2.$ 
Data from $e^+e^-$ and $ep$ are shown together with 
the MLLA predictions of
Dokshitzer {\it et al}~\protect\cite{dokevol}
(the full line is $Q_0=\Lambda,$ the dashed
$ Q_0= 2\Lambda,$ and the dotted $Q_0 = 3\Lambda$) and
the  limiting spectrum predictions of
Fong and Webber~\protect\cite{fongweb} (dash-dotted line where
available.)
The overlapping points are different $x$ ranges in the same $Q^2$ range.
The
inner error bars are the statistical error and the outer error bars are
the
systematic and statistical errors added in quadrature.}
\label{fig:qevol}
\end{figure}

The moments of the $\ln(1/x_p)$ distributions have been investigated
up to the 4th order~\cite{zeuslogxp};
the mean $(l),$ width $(w),$ skewness $(s)$ and kurtosis $(k)$
 were extracted from each distribution
by fitting a distorted Gaussian of the following form:
\begin{equation}
\frac{1}{\sigma_{tot}} \frac{d\sigma}{d\ln(1/x_p)} \propto
 \exp\left(\frac{1}{8}k-\frac{1}{2}s\delta -\frac{1}{4}(2+k)\delta^2
+\frac{1}{6}s\delta^3 + \frac{1}{24}k\delta^4\right), \label{eq:dg}
\end{equation}
where $\delta = (\ln(1/x_p) - l)/w,$
over a range of $\pm1.5$ units (for $Q^2 < 160 {\rm\ GeV^2}$) or 
$\pm2$ units (for $Q^2 \ge 160 {\rm\ GeV^2}$) in $\ln(1/x_p)$ around the
mean.
The equation
is motivated by the expression used for the MLLA predictions of the 
spectra~\cite{fongweb}.

Figure~\ref{fig:qevol} shows the moments of the  $\ln(1/x_p)$
spectra as a function of $Q^2.$
It is evident that the mean and width increase with
increasing $Q^2,$ whereas the skewness and kurtosis decrease. 
Similar fits performed on $e^+e^-$ data~\cite{logxpee} 
show a reasonable agreement with our results,
consistent with the universality of fragmentation for this distribution
at large $Q^2$.

The data are compared to the MLLA predictions of Ref.~\cite{dokevol}, 
using a value of $\Lambda=175{\rm \ MeV},$ 
for different values of $Q_0.$ A comparison is also made with
the predictions of Ref.~\cite{fongweb} for the
limiting spectrum  ($Q_0 = \Lambda$).
The MLLA predictions of the 
limiting spectrum in Ref.~\cite{dokevol}
describe the mean well. However both of the MLLA calculations predict a
negative skewness which tends towards zero
with increasing $Q^2$ in the case of the limiting spectra.
This is contrary to the measurements. The qualitative description of the
behaviour of the skewness with $Q^2$ can be achieved for a 
truncated cascade ($Q_0 > \Lambda$), but a consistent description of 
the mean, width, skewness and kurtosis
cannot be achieved. 

It can be concluded that the
MLLA predictions, assuming LPHD, do not describe the data.
It should be noted however that  a moments analysis has been 
performed~\cite{seroch}, taking into
account the limitations of the  massless assumptions of the MLLA
predictions;
this yields good agreement between the
limiting case of the MLLA~\cite{dokevol}
and ${\rm e^+e^-}$ data over a large energy range,
$ 3.0 < \sqrt{s_{ee}} < 133.0\ {\rm GeV}.$

\subsection{Evolution of the $\boldmath x_p$ Distributions} 
\label{sec:scaling}

Scaling violations are predicted in the fragmentation functions,
which represent the probability
for a parton to fragment into a particular
hadron carrying a given fraction of the parton's energy.
Fragmentation functions, like parton densities, 
cannot be calculated in perturbative QCD but
can be evolved with the hard-process scale,
using the DGLAP evolution~\cite{DGLAP} equations,
from a starting distribution at a defined energy scale; this starting
distribution can be derived from a fit to data.
If the fragmentation functions are combined with
the cross sections for the inclusive production of
each parton type in the given physical process, predictions can be made
for
scaling violations, expressed as the $Q^2$ evolution 
of the $x_p$ spectra of final state hadrons \cite{webnas}.
The NLO calculations (CYCLOPS)~\cite{dirk}
 of the scaled momentum distribution 
exist for DIS. 

\begin{figure}[!hbt]
\centerline{\epsfig{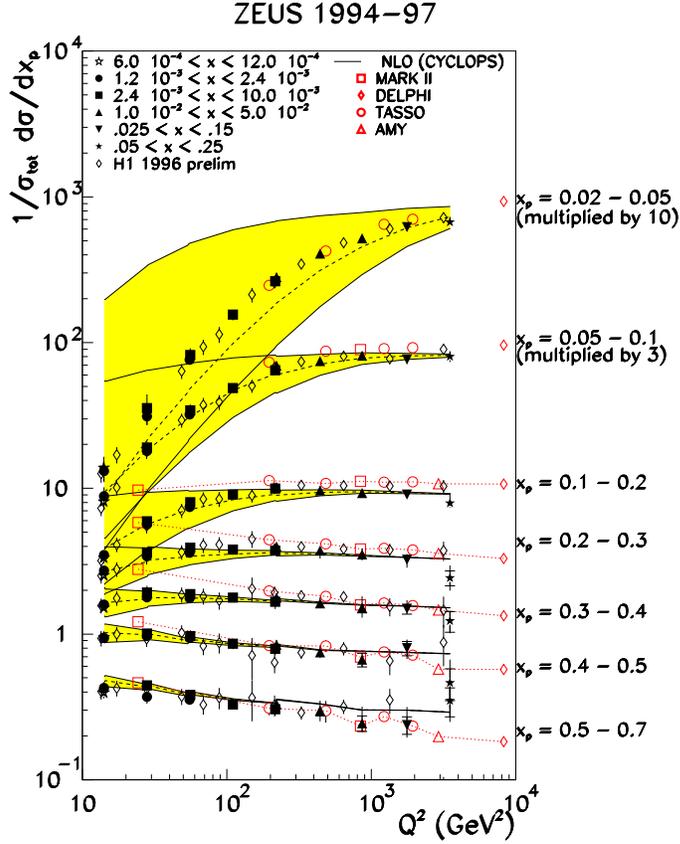}}
\caption{The inclusive charged particle distribution,
$ 1/\sigma_{tot}~ d\sigma/dx_p$,
in the current fragmentation region of the Breit frame.
The
inner error bar is the statistical and the outer error bar shows the
systematic and statistical errors added in quadrature.
The open points represent data from $e^+e^-$ experiments divided by two
to
account for $q$ and $\bar q$ production (also corrected
for contributions to the charged multiplicity from $K^0_S$ and $\Lambda$
decays). The low energy MARK II data has been offset slightly to the
left for the sake of clarity.
NLO predictions~\protect\cite{dirk} multiplied by the 
kinematic
correction described in the text. The shaded area represents the extreme
cases
$0.1  {\ \rm GeV} < m_{\rm eff} < 1.0  {\ \rm GeV}.$ The upper band
corresponds to $m_{\rm eff} = 0.1 {\ \rm GeV}$
 and the lower band $m_{\rm eff} = 1.0 {\ \rm GeV}.$}
\label{fig:largexp}
\end{figure}

The inclusive charged particle distribution,
$ 1/\sigma_{tot}~ d\sigma/dx_p$,
in the current fragmentation region of the Breit frame is shown in bins
of $x_p$ and $Q^2$ in Fig.~\ref{fig:largexp}.
The ZEUS~\cite{zeuslogxp} and H1~\cite{kant} data are in good agreement.
The fall-off as $Q^2$ increases for $x_p > 0.3$ 
(corresponding to  the production of more particles with
a smaller fractional momentum) is indicative of scaling
violations in the fragmentation function.
The distributions rise
with $Q^2$ for $x_p <0.1$ and are discussed in more detail below.
The data are compared
to $e^+e^-$ data~\cite{eedata} 
(divided by two to account for the production of a $q\bar q$ pair) 
at  $Q^2=s_{ee}.$
For the higher $Q^2$ values shown there is a good agreement between the
measurements in the
current region of the Breit frame in DIS and the $e^+e^-$
results; this again supports
the universality of fragmentation. 
The fall-off observed in the HERA data at low $x_p$ and low $Q^2$ 
is greater than that observed in $e^+e^-$ data at SPEAR~\cite{spear};
this can be attributed to
processes not present in $e^+e^-$ (e.g. scattering off a sea quark
and/or
boson gluon fusion (BGF)) which depopulate the current 
region~\cite{val,diffeflow}.

A kinematic correction has recently been suggested~\cite{durws}
to the NLO calculation~\cite{dirk}  of the inclusive charged particle
distribution which has the form,
$ 1/(1+(m_{\rm eff}/(Qx_p))^2),$
where $m_{\rm eff}$ is an effective
mass to account for the massless assumption used in the fragmentation
functions.
It is expected to lie in the range
$0.1 {\ \rm GeV} < m_{\rm eff} < 1.0  {\ \rm GeV}.$
The  $x_p$ data are compared to the CYCLOPS NLO QCD 
calculation incorporating this correction in
Fig.~\ref{fig:largexp}. This calculation
combines a full next-to-leading order matrix element
with the
${\rm MRSA^{\prime}}$ parton densities (with $\Lambda_{\rm QCD} =
230{\rm \ MeV})$
and NLO fragmentation functions
derived from fits to $e^+e^-$ data \cite{binnewies}.
The kinematic correction allows a more legitimate theoretical comparison
to lower $Q^2$ and $x_p$ than was possible in earlier 
publications~\cite{breit2}. The bands represent the uncertainty in
the predictions by taking the extreme cases of $m_{\rm eff}=0.1  {\ \rm
GeV}$
and  $m_{\rm eff}=1.0  {\ \rm GeV}.$ These uncertainties are large at
low
$Q^2$ and low $x_p,$ becoming smaller as $Q^2$ and $x_p$ increase.
Within these theoretical uncertainties there is good
agreement throughout the selected kinematic range. 
ZEUS found the kinematic correction describes the general trend of the data 
but it was not possible to achieve a good $\chi^2$ fit for $m_{\rm eff}$
over the whole $x_p$
and $Q^2$ range. In contrast H1 reported at DIS'99~\cite{kant} that a good
description of the data could be achieved with a value of $m_{\rm eff} = 0.6
{\ \rm GeV}.$ The dashed line shows the ZEUS NLO calculation from
CYCLOPS multiplied
by the kinematic correction using $m_{\rm eff} = 0.6 {\ \rm GeV}.$ There is
not a good description of the data, therefore it is concluded
the distribution also
depends strongly on the parameters used to generate the NLO predictions.
The uncertainties introduced by this kinematic correction
restrict to high $Q^2$ and high $x_p$ the kinematic range 
that may be used to extract $\alpha_s$ from the observed scaling
violations.

\section{Target Fragmentation}

\begin{figure}[htb!]
\begin{center}\mbox{\epsfig{file=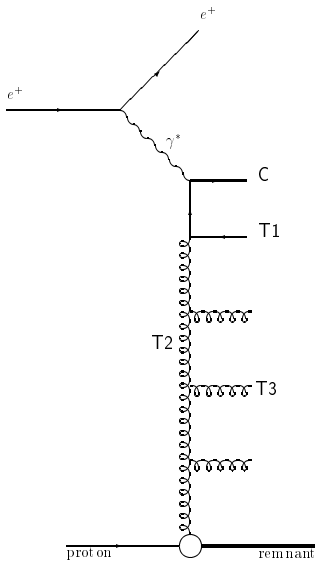,height=0.3\textheight}}
\end{center}
\caption{{ A schematic of
DIS scattering at low $x$ within the MLLA framework.
Quark C represents the struck sea 
quark in the current fragmentation region. T1 is
the other half of the quark box which is in the
target region. T2 is the t-channel
gluon exchange and T3 the rungs of the gluon ladder.}}
\label{pic1}
\end{figure}

DIS at low $x$ allows a study of fragmentation in
the target region following the initial scattering off a
sea quark (or antiquark). 
The description based on MLLA~\cite{dok} 
is shown schematically
in Fig.~\ref{pic1}, where the quark box at the top of the gluon
ladder represents the scattered sea quark plus its antiquark partner. 
The MLLA predictions are made up of a number of contributions.
Contribution C, the top leg of the quark box, corresponds to
fragmentation  of the outgoing quark
in the current region.
Three further contributions (T1, T2 and T3), which are sources of soft
gluons,
are considered in these analytical calculations to be associated
with the target region.
It is predicted~\cite{dok} that
the contribution T1( the bottom leg of the quark box)
behaves in the same way as the current quark C and so should have
 no $x$ dependence.  
The contribution T2 is due to the colour field
between the remnant and the struck quark,
and the contribution T3 corresponds to
the fragmentation of the rungs in the gluon ladder.
Both T2 and T3 are predicted to have $x$ and
$Q^2$ dependences which differ from T1.
Both the T1 and T2 contributions have been calculated and give
particles of momenta $<Q/2.$ 
The collinear gluons T3, on the other hand,
generally fragment to particles with momentum 
${\raisebox{-.6ex}{${\textstyle\stackrel{>}{\sim}}$}}Q/2.$ 
For values of the scaled momentum $x_p < 1.0,$
the region of phase space 
is analogous to the current region
and has contributions mainly from T1 and T2.
The parton momentum spectra
predicted by MLLA,  over a range of $Q^2$ and $x,$
are shown in more detail in Ref.~\cite{anis}.
In the target region these spectra are approximately 
Gaussian for  $ x_p < 1$;
they peak at a value of $x_p \sim 0.1-0.2$ in the range of $x$ and $Q^2$
measured by ZEUS~\cite{zeuslogxp}, falling to a plateau region 
for  $ 1 < x_p < (1-x)/x$ (the maximum value of $x_p$ in the target
region).

\begin{figure}[hbt!]
\begin{center}\mbox{\epsfig{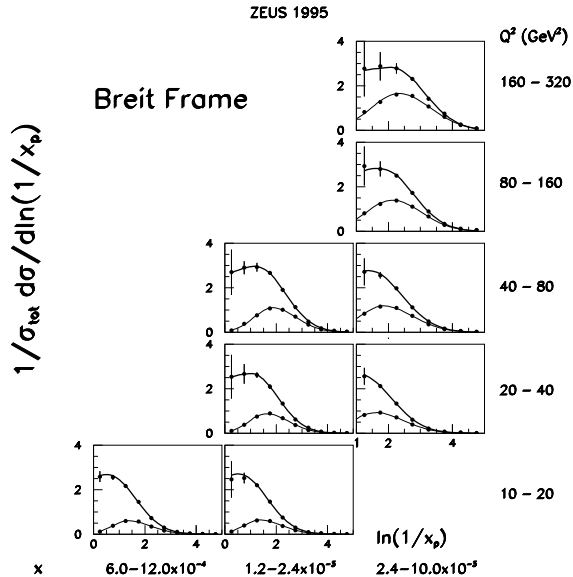}}
\end{center}
\caption{ The corrected $\ln(1/x_p)$ distributions  for the target
and current
regions for the 1995 data. Fitted two-piece normal distributions are
shown
to guide the eye. The heavy line
corresponds to
the target region, the light line to the current region.
The error bars are the sum of the statistical and systematic errors in
quadrature.}
\label{lxpda}
\end{figure}

The distributions in $\ln(1/x_p)$
are shown for both 
the target and current regions in Fig.~\ref{lxpda}. The
fitted curves shown are
two-piece normal distributions~\cite{boe} to guide the eye. 
In contrast to the current region,
the target region distribution does not fall
to zero as $\ln(1/x_p)$ tends to zero.
Although the magnitude of the single particle density at the
peak position of the current region distribution
grows by a factor
of about three over the $Q^2$ range shown, the single particle
density  of the target
distribution, at the $x_p$ value corresponding to the peak of the
current distribution
(contribution C is equivalent to contribution T1),
depends less strongly on $Q^2$ and 
increases by only about $30\%$.
In addition the $\ln(1/x_p)$ distribution 
shows no significant dependence on $x$ when $Q^2$ is kept constant.
In the target region the peak position of the $\ln(1/x_p)$ distribution
increases more rapidly with $Q^2$ than in the current region; this is
consistent with the behaviour expected from cylindrical phase space.
The approximate Gaussian distribution of the MLLA predictions peaking
at $\ln(1/x_p) \sim 1.5-2.5$~\cite{anis} is not observed. The results
strongly suggest that the target
distributions are inconsistent with the MLLA predictions  when used
in conjunction with LPHD.

\section{Rapidity Distributions}

 There are predictions, based on LPHD, for the rapidity 
distribution of
charged particles in the Breit frame~\cite{ochs}. (Rapidity is defined as
$Y= \ln\left((E+p_z)/(E-p_z)\right),$ where $E$ is the energy of the
particle and $p_z$ is the longitudinal component of its momentum.)
It is predicted that there is a sharp
rise in the charged particle density
 followed by a plateau with a width proportional to  $\ln(Q)$ as
one moves from the current to the target fragmentation region.
It is also predicted that there will be another increase in the particle
density and an appearance
of a second plateau, with the ratio  of the two plateaux being
$9/4.$ This ratio reflects the change in colour charge from a dominant
quark one, in the current region, to the dominant
gluonic one in the target fragmentation region.

\begin{figure}[phbt!]
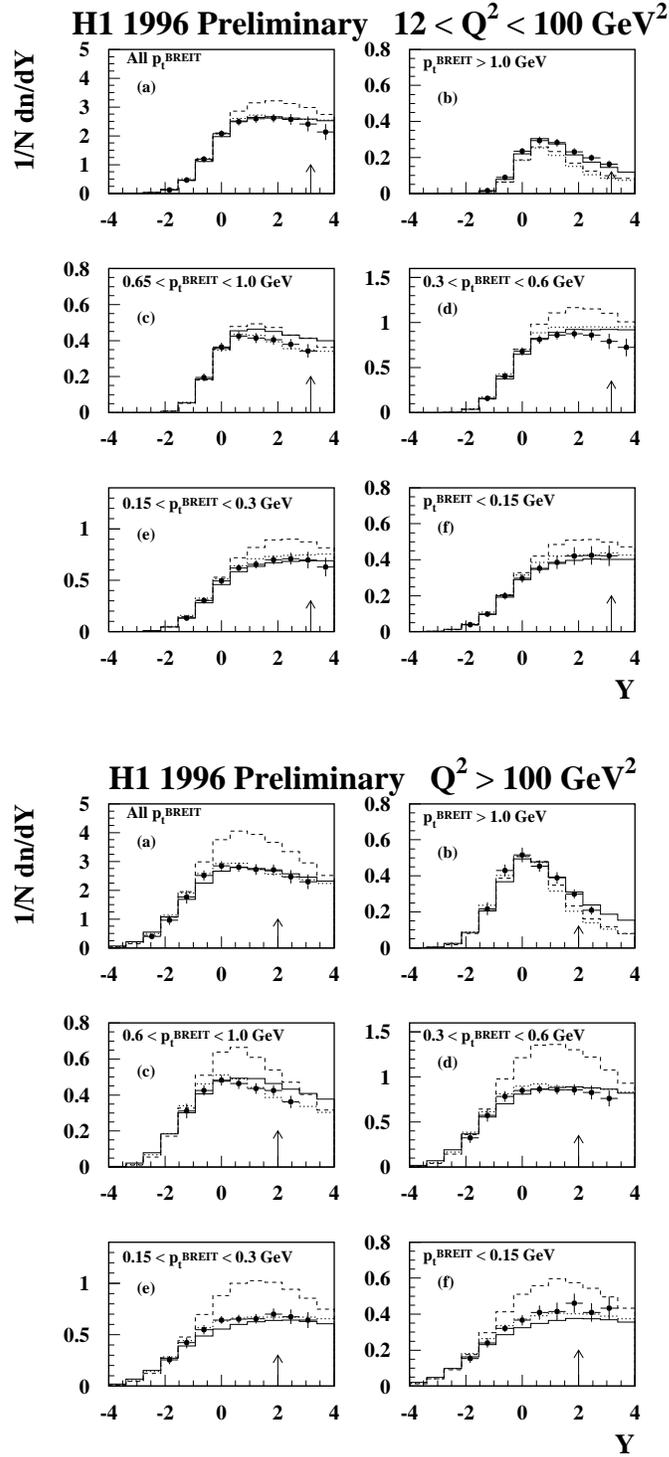

\begin{center}\epsfig{file=brook.4a,height=0.4\textheight}
\epsfig{file=brook.4b,height=0.4\textheight}
 \end{center}
\caption{  The rapidity distributions of charged tracks in intervals of
transverse momentum and $Q^2$ in the Breit frame. The error bars are the
sum of the statistical and systematic errors added in quadrature. The
arrow indicates the position of the origin of the hadronic centre of
mass system for the $\langle Q \rangle$ of the data. The histograms show
the predictions of LO Monte Carlo models, the solid line is the
ARIADNE~\protect\cite{ARIADNE} Monte Carlo and the 
dashed (dotted) line the LEPTO ME+PS~\protect\cite{LEPTO}
prediction with (without) soft colour interactions.
}
\label{kant}
\end{figure}

The rapidity distributions for charged particles are shown in
Fig.~\ref{kant}~\cite{kant}. 
A flat plateau is observed at low transverse momentum,
$p_t.$ As $p_t$ increases , QCD effects gradually evolve the flat plateau
into an approximate Gaussian, peaking near zero. This illustrates the
nature of the Breit frame in separating the current and target fragmentation
region.
The expected step between the current and target region of the
rapidity spectra is not observed.
Also shown in Fig.~\ref{kant} is a comparsion of the data with
predictions from LO Monte Carlo models. In general the data is well
described. ARIADNE~\cite{ARIADNE} agrees well with the data
but LEPTO~\cite{LEPTO} has problems 
in describing the high $p_t$ data and the
introduction of soft colour interactions~\cite{SCI} 
destroys the agreement with the data. 

\section{Current-Target Correlations}

The correlation coefficient $\kappa$: 
\begin{equation} 
\kappa =\sigma_c^{-1} \sigma_t^{-1}\mathrm{cov}(n_c , n_t)
\qquad
\mathrm{cov}(n_c , n_t) = \langle n_c n_t \rangle -  \langle n_c \rangle \langle n_t \rangle,
\label{1c}
\end{equation}
is used to measure the dependence between
charged particle production in the current region, $n_c,$
and production in the target region, $n_t,$
where $\sigma_c$ and $\sigma_t$ are  the standard deviations of 
the multiplicity distributions 
in the current and target regions respectively. 
For positive correlations, $\kappa$  is  positive whilst for
anti-correlations it is  negative.
At low $Q^2$ these correlations are sensitive to the BGF
process which depends on the gluon density of the proton~\cite{sergei}. 

\begin{figure}
\begin{center}
\epsfig{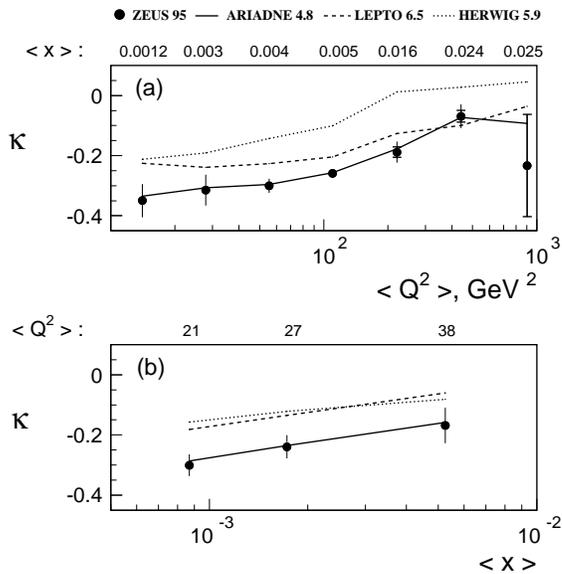}
\caption{ 
{\bf (a)} represents the  evolution of  the 
coefficient of correlations $\kappa$
with predominant variation in  $Q^2$
for corrected data and MC predictions; 
{\bf (b)} shows the same quantity where predominantly $x$ varies. 
The  corrected values of  
$\langle Q^2\rangle$ and $\langle x \rangle$ are indicated for each plot.    
The inner error bars on the data show the statistical uncertainties.  
The full error bars  include the systematic uncertainties. } 
\label{fig:rho}
\end{center}
\end{figure}

Figure~\ref{fig:rho}
shows the behaviour of  the correlation coefficient
$\kappa$ as a function of the average values of $Q^2$ and  $x$. 
Anti-correlations are  observed for all values 
of $x$ and $Q^2$~\cite{zeuscorrel}.
The  magnitude  of $\kappa$ 
decreases  with increasing  $\langle Q^2 \rangle $
from $0.35$ to $0.1$.  
According to the analytic results of~\cite{sergei} 
these observed anti-correlations  can be due to   
the ${\cal O}(\alpha_s )$ effects (QCDC and BGF). The ${\cal O}(\alpha_s
)$ kinematics
in the Breit frame can reduce the particle multiplicity  in the
current region and increase it in the target region.   
The magnitude of the anti-correlations  increases
with decreasing $\langle x \rangle$. 
According to ~\cite{sergei}
this can be due to an increase of the fraction of events  with one or 
two jets produced in the target region. This behaviour is driven by 
an increase of the BGF rate, due to
an increase in the gluon density inside the proton.
These observations are, qualitatively, consistent with the depopulation
effects discussed in section~\ref{sec:scaling}.

In addition, Fig.~\ref{fig:rho} shows a comparison of the data with
various LO Monte Carlo models.
The ARIADNE model agrees  well with the data.
The LEPTO and HERWIG~\cite{HERWIG} predictions show  
the same trend as the data but
do not reproduce the magnitude of the correlations.

\section{Event Shapes}

The event shape dependence on $Q$
can be due to the logarithmic change of the strong coupling constant
$\alpha_s(Q) \propto 1/\ln Q$, and/or
power corrections (hadronisation effects)
which are expected to behave like $1/Q$.
Recent theoretical developments suggest that $1/Q$ corrections are not
necessarily related to hadronisation, but may instead be a universal
soft
gluon phenomenon associated with the behaviour of the running coupling
at
small momentum scales~\cite{webdok}.
These non-perturbative corrections are governed by a parameter
$\bar\alpha_{0}.$
The scale dependence of any event shape mean $\langle F \rangle$ can
written as the sum of two terms: one associated with the  perturbative 
contribution, $\langle F \rangle^{\rm pert},$
and the other related to the power corrections, $\langle F \rangle^{\rm
pow}.$ The perturbative contribution to an event shape can be calculated from
NLO programs, such as DISENT~\cite{DISENT}.

At HERA a number of infrared-safe event-shape variables have been
investigated~\cite{rabbertz}. Their definitions are given below,
where the sums extend over all hadrons $h$
(being a calorimetric cluster in the detector or a parton in the QCD
calculations)
with four-momentum
$p_h = \{ E_h,\, {\bf p}_h \}$  in the current hemisphere of the
Breit frame.
The current hemisphere axis ${\bf n}  = \{0,\,0,\,-1 \}$
coincides with the virtual boson direction.
\begin{itemize}
  \item {\bf Thrust {\boldmath $T_c$}}
    \begin{eqnarray}
      T_c & = & \max \, \frac{\sum_h |\, {\bf p}_h\cdot {\bf n}_T \,
|}
                {\sum_h |\, {\bf p}_h \, |}
      \quad\quad\quad\quad\quad\quad\quad\quad \
      {\bf n}_T \ \equiv \ \mbox{thrust axis} \ ,
      \quad \ \
      \nonumber
    \end{eqnarray}
  \item {\bf Thrust {\boldmath $T_z$}}
    \begin{eqnarray}
      T_z & = & \frac{\sum_h |\, {\bf p}_h\cdot {\bf n} \, |}
                 {\sum_h |\, {\bf p}_h \, |}
          \ = \ \frac{\sum_h |\, {\bf p}_{z\,h}\, |}
                     {\sum_h |\, {\bf p}_h \, |}
      \quad\quad\quad\quad \ \ \
      {\bf n} \ \equiv  \ \mbox{hemisphere axis} \ ,
      \nonumber
    \end{eqnarray}
  \item {\bf Jet Broadening {\boldmath $B_c$}}
    \begin{eqnarray}
        B_c & = & \frac{\sum_h |\, {\bf p}_h\times {\bf n} \, |}
                     {2\,\sum_h |\, {\bf p}_h \, |}
          \ = \ \frac{\sum_h |\, {\bf p}_{\perp\,h}\, |}
                     {2\,\sum_h |\, {\bf p}_h \, |}
      \quad\quad\quad\quad
      {\bf n} \ \equiv \ \mbox{hemisphere axis} \ ,
      \nonumber
    \end{eqnarray}
  \item {\bf Scaled Jet Mass {\boldmath $\rho_c$}}
    \begin{eqnarray}
      \rho_c & = & \frac{M^2}{Q^2}
      \ = \ \frac{(\, \sum_h \, p_h \, )^2}{Q^2} \ .
      \phantom{xxxxxxxxxxxxxxxxxxxxxxxxxxxxxxx}
      \nonumber
    \end{eqnarray}
  \item {\bf {\boldmath $C$}  Parameter }
    $$
      C  =  3(\lambda_1\lambda_2 + \lambda_2\lambda_3 +
\lambda_3\lambda_1) 
    $$
with $\lambda_i$ being the eigenvalues of the momentum tensor
    $$
      \Theta_{jk}   =  \frac{\sum_h
\frac{p_{j_h}p_{k_h}}{|{\bf p}_h|}}{\sum_h |{\bf p}_h|} 
    $$
\end{itemize}
Also investigated was the variable
$y_{fJ}$ (over the whole of phase space).
This variable represents the transition value for
$(2+1) \rightarrow (1+1)$ jets of the factorizable JADE  jet
algorithm for a particular event.

\begin{figure}[htb]
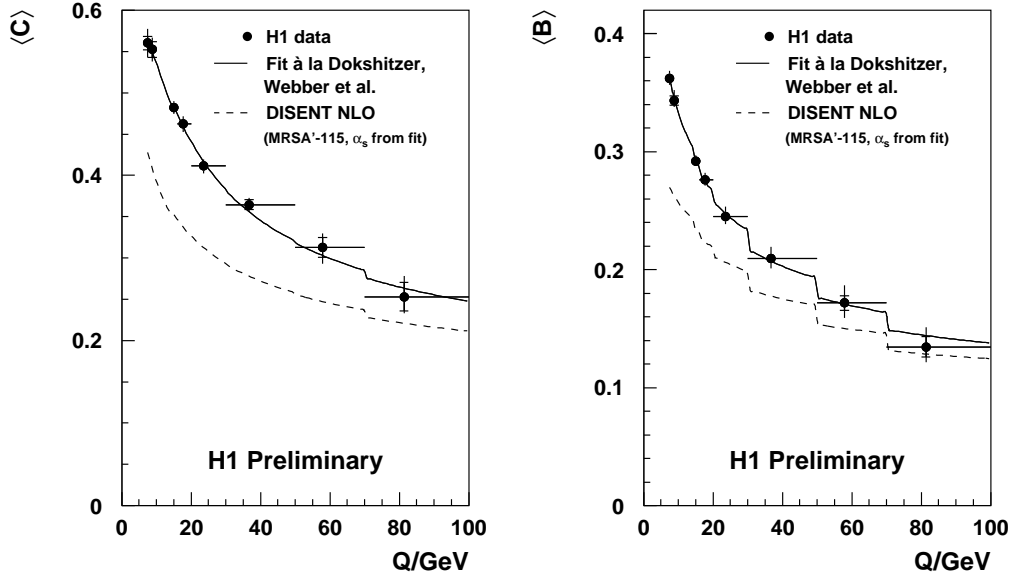

\begin{center}
\epsfig{file=brook.6a,width=0.4\textwidth,bbllx=0pt,bblly=0pt,bburx=220pt,bbury=260pt}
\epsfig{file=brook.6b,width=0.4\textwidth,bbllx=0pt,bblly=0pt,bburx=220pt,bbury=260pt}
\caption{The points are the corrected mean values of $C$ and $B$ as a
function of $Q.$ 
The inner error bars on the data show the statistical uncertainties.  
The full error bars  include the systematic uncertainties. The full line
corresponds to a power correction fit according to the approach
in~\protect\cite{webdok}. The dashed line is the perturbative (NLO)
predction from DISENT ising the value of $\alpha_s$ found from the full
fit.} 
\label{meandata}
\end{center}
\end{figure}

A common feature of all the mean event shape values
is the fact that they exhibit a decrease with rising $Q$,
Fig.~\ref{meandata}.
This is due to fact
that the energy flow becomes more collimated along the event
shape axis as $Q$ increases, a phenomenon also observed in $e^+e^-$
annihilation experiments.

A simple ansatz for the power correction would be $\langle F
\rangle^{\rm pow} = \Gamma/Q.$ 
However the fits using $\Gamma$ alone are
poor and support the more detailed approach outlined in~\cite{webdok}.
In this approach $\langle F \rangle^{\rm pow}$ is parameterised as follows:

\begin{eqnarray}
\langle F \rangle^{\rm pow} &  = &  a_F\frac{32}{3\pi^2}{\cal M}
\left(\frac{\mu_I}{Q}\right) \nonumber \\
 & &   \left[\bar\alpha_{0}(\mu_I) - \alpha_s(Q) -
\frac{\beta_0}{2\pi}\left(\ln\frac{Q}{\mu_I}+\frac{K}{\beta_0}+1\right)\alpha_s^2(Q)\right]
\end{eqnarray}

\noindent where $\beta_0,K$ are constants dependent on the number of flavours,
$a_F$ is a calculable coefficient dependent on the observable $F$,
$\mu_I$ is an `infra-red' matching scale $(\mu_I= 2{\rm\ GeV}),$
$\frac{2}{\pi}{\cal M} \approx 1.14$ is a 2-loop correction (known as
the Milano factor) and $\bar\alpha_{0}$ is an universal,
non-perturbative effective strong coupling below $\mu_I.$

\begin{figure}[hbt]
\begin{center}
\epsfig{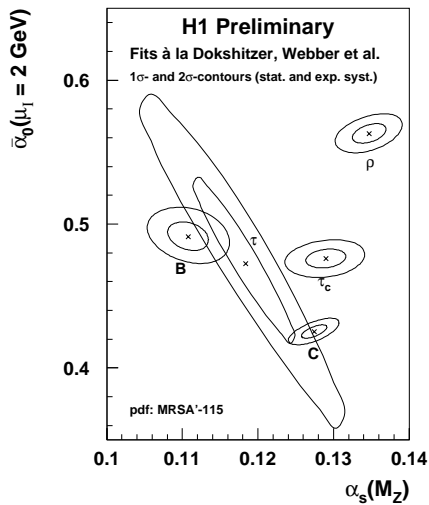} 
\caption{Results of the fit to $\bar\alpha_0,\alpha_s$ for the means of
$\tau,B,\tau_C,\rho {\rm\ and\ } C.$ The ellipses illustrate the
$1\sigma$ and $2\sigma$ contours including both statistical and
systematic uncertainties.}
\end{center}
\label{ref:fit}
\end{figure}

The results of the fit are shown in Fig.~\ref{ref:fit}. The parameter
$\bar\alpha_0$ is observed to be 
$\approx 0.5$ for all event shapes (except the jet rate parameter
$y_{fJ}$) , consistent with theoretical expectation. 
However there is a large spread in the values of $\alpha_s.$ 
The theoretically calculated parameter for the power corrections
for $y_{fJ}$ was $a_{y_{fJ}} = 1$. This is contrary to the
observed need for small negative hadronisation corrections~\cite{rabbertz}. 
A reasonable
fit for $y_{fJ}$ was achieved by using a value of $a_{y_{fJ}} = -0.25.$
The extended analysis of the mean event shapes in DIS are consistent with
the application of power corrections according to~\cite{webdok} though
there is still need for further understanding.

\section{Summary}
To understand the underlying QCD processes in DIS
 it is necessary to study the
hadronic final state.
At the current level of understanding, QCD works well and
describes the HERA data.
As the precision of the HERA data improves and the NLO QCD calculations
become available the framework of QCD is being tested more thoroughly.

%

\end{document}